\documentstyle[12pt,psfig]{article}

\def\baselinestretch{1.2}
\def\href#1#2{#2}  

\newcommand{\norm}[1]{\raise.3ex\hbox{:} #1 \raise.3ex\hbox{:}\,}

\newcommand{\beq}{\begin{equation}}
\newcommand{\eeq}{\end{equation}}

\def\appendix{{\newpage\section*{Appendix}}\let\appendix\section%
        {\setcounter{section}{0}
        \gdef\thesection{\Alph{section}}}\section}

\begin{document}

\begin{titlepage}

\begin{flushright}
NSF-ITP-99-020\\
hep-th/9903227
\end{flushright}
\vfil\vfil

\begin{center}

{\Large {\bf Holographic description of D3-branes in flat space}}

\vfil

Akikazu Hashimoto\\

\vfil

Institute for Theoretical Physics\\ University of California,
Santa Barbara, CA 93106\\
aki@itp.ucsb.edu\\

\vfil
\end{center}

\begin{abstract}
\noindent 
We describe a scheme for constructing the holographic dual of the full
D3-brane geometry with charge $K$ by embedding it into a large anti-de
Sitter space of size $N$.  Such a geometry is realized in a
multi-center anti-de Sitter geometry which admits a simple field
theory interpretation as $SU(N+K)$ gauge theory broken to $SU(N)
\times SU(K)$.  We find that the characteristic size of the D3-brane
geometry is of order $(K/N)^{1/4} U^0$ where $U^0$ is the scale of the
Higgs. By choosing $N$ to be much larger than $K$, the scale of the
D3-brane metric can be well separated from the Higgs scale in the
radial coordinate.  We generalize the holographic energy-distance
relation and estimate the characteristic energy scale associated with
these radial scales, and find that the $E/U$ relation becomes
effectively $U$ independent in the range $(K/N)^{1/2} U^0 < U <
U^0$. This implies that all detailed structure of the D3-brane
geometry is encoded in the fine structure of the boundary gauge theory
at around the Higgs scale.
\end{abstract}

\vfil\vfil\vfil
\begin{flushleft}
March 1999
\end{flushleft}

\end{titlepage}
\renewcommand{\baselinestretch}{1.05}  

The conjecture of Maldacena \cite{adscft} relating type IIB string
theory on $AdS_5 \times S_5$ to ${\cal N}=4$ SYM on 3+1 dimensions was
originally formulated by taking the ``near horizon'' or the
``decoupling'' limit of D3-branes sitting in flat space. Even before
the formulation of the Maldacena conjecture, the close connection
between the gauge theory living on the world volume of a D-brane and
the supergravity background was explored in the context of computing
absorption cross-sections \cite{absorb1,absorb2,absorb3}.  Once the
details of the correspondence between the bulk and the boundary
theories were better understood \cite{gkp,wittenads}, it became
sensible to discard the asymptotically flat region of the D3-brane
background and work entirely with the near horizon anti-de Sitter
geometry. The reason is that the Maldacena conjecture can be stated
concretely in a form of an identity of suitably defined generating
functions between string theory on anti-de Sitter space and the gauge
theory on its boundary. One speaks of the boundary theory being
``holographically dual'' to the bulk theory
\cite{thooftholo,sussholo}, and in the case of $AdS_5\times S_5$
arising from the near horizon limit of the D3-brane, both the
bulk\footnote{This is true up to the subtle issue of defining string
theories in a presence of RR background. Currently, only the
supergravity approximation which can be trusted at low energies is
properly formulated.} and the boundary theory are well defined.

Just because we have come to understand the holographic dual of the
near horizon $AdS_5 \times S_5$ geometry does not mean that the
holographic dual of the full D3-brane geometry does not exist. Indeed,
we expect all theory of gravity to have a holographic dual
\cite{thooftholo,sussholo}. It is therefore quite natural to ask what
the boundary dual to string theory in the full D3-brane background
might be.  In this note we provide an answer to this question,
although in an implicit form.

It is easy to appreciate the scope of this problem: the holographic
dual on the boundary is likely to be a very complicated theory. The
D3-brane background and the $AdS_5 \times S_5$ background behave more
or less the same way in the small radius region, but differ
drastically at large radius. In holography, small and large radius
regions of the bulk theory are associated with the infra-red and
ultra-violet of the boundary theory, respectively
\cite{adscft,susswitt}. This means that the UV structure of the
boundary theory must be drastically different than that of the SYM
theory.  In \cite{GHKK}, it was suggested that this theory might be
realized as a condensate of the irrelevant operator $F^4$. However,
without the detailed knowledge of the UV structure of the theory, it
is not clear if a sensible meaning can be attached to condensation of
an irrelevant operator (see e.g.~\cite{wittenads}). Since the geometry
of the D3-brane metric is asymptotically flat, it is quite likely that
the boundary theory, whatever it may be, is as complicated as the
holographic dual of the full type IIB string theory on Minkowski space
\cite{witstr98,joeS,sussflat}.

In this note, we resolve the ambiguity of the UV structure of the
boundary theory by embedding it in some other theory whose UV
structure is better understood.  The boundary theory of interest to us
can be extracted by flowing along the renormalization group down to
the appropriate scale. Although such a procedure will always lead to a
UV complete description of an effective field theory, it is generally
difficult to determine what UV fixed point can give rise to a
renormalization group flow that reaches the effective theory of
interest. In a holographic theory, however, it is very straight
forward to embed an arbitrary IR structure of interest into a well
defined UV structure. The idea is to simply embed the bulk geometry of
the IR theory somewhere in the near core region which smoothly
interpolates to the bulk geometry of the UV fixed point.  A very
convenient bulk geometry to use as a UV fixed point is the anti-de
Sitter space, since we understand the UV structure from the boundary
point of view as well\footnote{Geometric embeddings of an IR dynamics
in some UV theory utilizing the holographic principle was discussed in
\cite{IMSY,rg1,rg2,rg3}.}. In order to formulate the holographic dual
of the D3-brane geometry, all that we have to do is to consider a bulk
geometry which at small radius looks more or less like a D3-brane
geometry, but behaves at very large radius like an anti-de Sitter
space.

It turns out that a supergravity solution which interpolates between a
full D3-brane metric (with the asymptotically flat region) and an
anti-de Sitter space is very easy to construct. One needs to look no
further than the multi-centered anti-de Sitter solution. Let us
demonstrate this explicitly. Consider a type IIB supergravity solution
for parallel D3-branes with charges $N \gg K \gg 1$.
\beq ds^2 = \left(1 + {4 \pi g N \alpha'^2 \over (\vec{x}_\perp +
\vec{x}^0_\perp)^4} + {4 \pi g K \alpha'^2 \over {\vec{x}_\perp}^4}
\right)^{-1/2} d\vec{x}_\parallel^2 + \left(1 + {4 \pi g N \alpha'^2
\over (\vec{x}_\perp + \vec{x}^0_\perp)^4} + {4 \pi g K \alpha'^2
\over {\vec{x}_\perp}^4} \right)^{1/2} d\vec{x}_\perp^2. \eeq
Taking the decoupling limit keeping $\vec{U}_\perp = \vec{x}_\perp / \alpha'$ fixed gives
\beq ds^2 = \alpha' \left({4 \pi g N \over (\vec{U}_\perp +
\vec{U}^0_\perp)^4} + {4 \pi g K  \over {\vec{U}_\perp}^4}
\right)^{-1/2} d\vec{x}_\parallel^2 + \alpha' \left({4 \pi g N
\over (\vec{U}_\perp + \vec{U}^0_\perp)^4} + {4 \pi g K
\over {\vec{U}_\perp}^4} \right)^{1/2} d\vec{U}_\perp^2. \label{metric2}\eeq
If we restrict to range of parameters $|\vec{U}| \approx  (k/N)^{1/4}
|\vec{U}^0_\perp| \ll |\vec{U}^0_\perp|$, the metric simplifies to
\beq ds^2 = \alpha' {|{\vec{U}^0_\perp}|^2 \over \sqrt{4 \pi g
N}}\left(1 + {(K/N) |\vec{U}_\perp^0|^4 \over |\vec{U}|^4}
\right)^{-1/2} d\vec{U}_\parallel^2 + \alpha' { \sqrt{4 \pi g N} \over
|{\vec{U}^0_\perp}|^2} \left(1 + {(K/N) |\vec{U}_\perp^0|^4
\over |\vec{U}|^4} \right)^{1/2} d\vec{U}_\perp^2. \eeq
Finally, a simple  rescaling 
\beq \vec{y}_\parallel  =  { \sqrt{\alpha'} |{\vec{U}^0_\perp}| \over (4 \pi g
N)^{1/4}} \vec{x}_\parallel, \quad
\vec{r}_\perp =  { (4 \pi g N \alpha'^2)^{1/4} \over
|{\vec{U}^0_\perp}|} \vec{U}_\perp\eeq
leads to the standard from of the D3-brane metric
\beq ds^2 = \left( 1 + {4 \pi g K \alpha'^2 \over r^4}\right)^{-1/2}
d\vec{y}_\parallel^2 + \left( 1 + {4 \pi g K \alpha'^2 \over
r^4}\right)^{1/2} d \vec{r}_\perp^2. \eeq
Essentially, the $4 \pi g N /(\vec{U}_\perp+\vec{U}_\perp^0)^4$ term
in the harmonic function at the vicinity of $U = 0$ acts as the ``1''
to simulate the asymptotically flat region. An important point that we
learn from this exercise is that the characteristic length scale
$(K/N)^{1/4} |\vec{U}_\perp^0|$ associated with the D3-brane metric is
clearly well separated from the Higgs scale $|\vec{U}^\perp_0|$ or the
scale of the cosmological curvature $(4 \pi g N)^{1/4} /
\alpha'^{1/2}$. Therefore, there is plenty of room to fit an entire
D3-brane metric with an asymptotically flat region, especially if we
take the large $N$ limit while focusing on scales of order
$(K/N)^{1/4} |\vec{U}_\perp^0|$. This separation of scales was
possible because we chose to scale $K$ and $N$ differently.

In a certain sense, the point being made here is rather trivial in
light of the fact that one takes a large $N$ limit of the anti-de
Sitter space with RR charge $N$ to recover the flat space
\cite{joeS,sussflat}. If we combine such a flat space limit with a
source for $K$ units of RR flux, we were bound to recover the full
D3-brane metric.

The interpretation of multi-centered anti-de Sitter solution as the
SYM with Higgs VEV of order $\vec{U}_\perp^0$ have been discussed by
many authors \cite{adscft,DougTayl,YYWu}. One surprising aspect of SYM
with large 't Hooft coupling is the fact that energies associated with
this VEV is not $|\vec{U}_\perp^0|$ but $|\vec{U}_\perp^0|/\sqrt{gN}$
\cite{PeetPol}. In other words, gauge theory observable such as the
two point function $\langle {\cal O}(x) {\cal O}(0) \rangle$ will
only feel the effect of Higgsing when $1/x$ becomes of order
$|\vec{U}_\perp^0|/\sqrt{gN}$. Having identified $(K/N)^{1/4}
|\vec{U}^0_\perp|$ as an important scale in describing the D3-brane
geometry, we would like to know to what energy this scale corresponds
from the point of view of the boundary theory. We will address this
issue in the remainder of this note.

In order to associate an energy scale to $(K/N)^{1/4}
|\vec{U}_\perp^0|$, we need to generalize the notion of
``energy-distance'' relation
\beq E = {U^{(5-p)/2} \over g_{YM} \sqrt{N}}, \label{eurel} \eeq
derived in \cite{susswitt,PeetPol}, to multi-center backgrounds.  This
formula was derived in \cite{PeetPol} by computing the ``skin depth''
of the boundary fluctuation into the bulk, and by utilizing the
holographic information bound in \cite{susswitt}. It turns out,
however, that this same formula can be derived in many different
ways.\footnote{We thank Sunny Itzhaki for many discussions on this
point.} We will use the method most convenient for us, but let us list
a few example of an alternative method for deriving (\ref{eurel}).

One such approach is to recall the calculation of Wilson loop
expectation value as minimal surfaces \cite{JuanWilson,BISY}. When a
string is suspended from infinity with its endpoint separated by  $L = 1/E$, it
penetrates the radial direction down to some radius $U$. The relation
between $U$ and $L = 1/E$ is equivalent to (\ref{eurel})\footnote{See
equation (5.2) of \cite{BISY}.}.  Perhaps the methods developed in
\cite{MinWarn} can be adopted for this purpose.

An alternative approach is to consider a light-cone originating from a
fixed point in the boundary \cite{GarySunny}.  In a given time $t$,
the signal on the boundary will spread in size by the rate determined
by the speed of light. The energy associated with this size is
therefore $E = 1/t$. In the same time interval, the light signal will
travel in the radial direction by some amount $U(t)$. The relation
between $U$ and $E$ determined this way also leads to (\ref{eurel}).

For our application, this last approach is very convenient since all that is required is to solve the null geodesic equation
\beq
dt = \left({4 \pi g N
\over (U + U^0)^4} + {4 \pi g K
\over {U}^4} \right)^{1/2} d U \label{nullgeod},
\eeq
where for convenience, we have oriented $\vec{U}$ to be parallel to
$\vec{U}^0$ and dropped the vector sign. Choosing different
orientations will not drastically change the story as long as the
signal do not fall into the region near the $N$ branes.  Note that
if we set $U^0=0$ and $K=0$, this equation can be integrated trivially
to give
\beq t = {\sqrt{4 \pi g N} \over U} = {1 \over E} \label{ubig}, \eeq
which is equivalent to (\ref{eurel}). If $U^0$ and $K$ are both
non-vanishing, we will have to work a little harder. To get a rough
estimate of $t(U)$, it is convenient to divide the parameter space
for $U$ into regions: (a) $U^0 < U$, (b) $(K/N)^{1/4}U^0 < U < U^0$, and
(c) $U < (K/N)^{1/4}U^0$. In region (a), we might as well set $U^0 = 0$
and we get the essentially same result as (\ref{ubig})
\beq t(U) = {\sqrt{4 \pi g (N+K)} \over U} \approx {\sqrt{4 \pi g N}
\over U}.  \label{rega} \eeq
In region (b), (\ref{nullgeod}) can be approximated by
\beq dt = {\sqrt{4 \pi g N} \over (U^0)^2} dU, \eeq
so we find
\beq t(U) = {\sqrt{ 4 \pi g N} \over (U^0)^2} (U^0 - U) +t(U^0)
= {\sqrt{ 4 \pi g N} \over U^0} \left(2 - {U\over U^0} \right)
\label{regb}.  \eeq
Finally, in region (c), (\ref{nullgeod}) can be approximated by
\beq dt = {\sqrt{4 \pi g K} \over U^2} dU, \eeq
so we find
\beq t(U) = \sqrt{4 \pi g K} \left({1\over U} - {1 \over (K/N)^{1/4}
U^0} \right) + t((K/N)^{1/4}U^0) = {\sqrt{4 \pi g K} \over U} +
{\sqrt{ 4 \pi g N} \over U^0} \left(2 - 2 (K/N)^{1/4} \right)
\label{regc}. \eeq

\begin{figure}
\centerline{\psfig{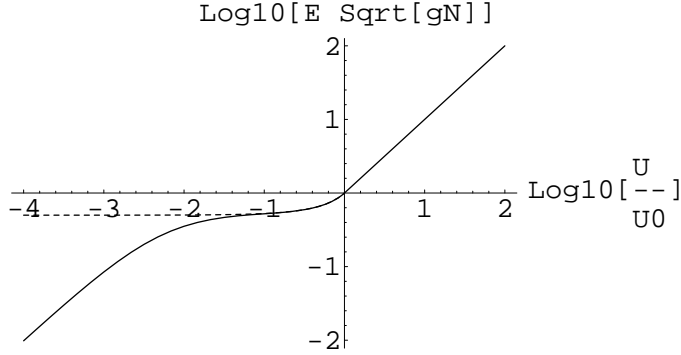}}
\caption{Log-Log plot of the $E/U$ function for the multi-centered
$AdS_5 \times S_5$.  The solid line is for $K/N = 10^{-4}$. The dashed
line is for $K/N=0$.\label{figa}}
\end{figure}

Inverting (\ref{rega}), (\ref{regb}), and (\ref{regc}), we seem to be
finding that in region (a),
\beq E(U) \approx {U \over \sqrt{4 \pi gN}} \label{eua}\eeq
In region (b) by the time $U$ reaches $(K/N)^{1/4} U^0$, $E(U)$ is given by
\beq E(U)  \approx {1 \over 2} {U^0 \over \sqrt{4 \pi gN}}
\left(1 + {1 \over 2} \left({K \over N}\right)^{1/4}\right), \eeq
which is of the same order of magnitude as $E(U^0)$. In fact, $E(U)$
will continue to be in the same order of magnitude even in region (c)
until $U$ become smaller than $\sqrt{K/N} U^0$ at which point the
first term in (\ref{regc}) starts to dominate, leading to
\beq E(U) = {U \over \sqrt{4 \pi g K}}. \label{euc} \eeq
Since we do not rely on this kind of analysis beyond estimating orders
of magnitudes, the picture that seems to be emerging is the following:
the ``energy-distance'' relation for $U > U^0$ behaves like
(\ref{eua}), while for $U < \sqrt{K/N} U^0$ it behaves like
(\ref{euc}). For the values of $U$ in between these regions, $E(U)$
will remain more or less constant at the energy of order Higgs scale
$E = U/\sqrt{gN}$. Note that the scale of interest to us, namely $U
\approx (K/N)^{1/4} U^0$ is precisely in this flat region. See figure
\ref{figa} for an illustration.

The multi-center solution breaks the $S_5$ spherical symmetry, but
this symmetry can be restored by smearing the supergravity solution
(\ref{metric2}) along the $S_5$. This leads to a geometry of the form
\beq ds^2 = \alpha' f(U)^{-1/2} d\vec{x}_\parallel^2 + \alpha'
f(U)^{1/2} d\vec{U}^2. \eeq
where
\beq
f(U) = \left\{ 
\begin{array}{ll}
{4 \pi g (N+K) \over U^4} & (U^0 < U) \\
{4 \pi g N \over (U^0)^4} \left(1+ {K (U^0)^4 \over N U^4}\right) & (U <  U^0) 
\end{array}
\right.
\eeq
which is that of an $AdS_5 \times S_5$ for $U > U^0$ and a full
D3-brane geometry of size $(K/N)^{1/4} U^0$ for $U < U^0$.  If we set
$K = 0$, this is precisely the flat space region described in
\cite{KLT}.  Turning on $K$ then has a simple interpretation as
placing a small D3-brane in this flat space-time patch. One can apply
the method of \cite{gkp,wittenads} for computing the correlation
function of chiral primary operators by solving the wave equation for
modes propagating in this background. At linear order in fluctuations,
the minimal scalar equation will be of the standard Bessel form for
$U^0 < U$, whereas for $U < U^0$, it will be of the Mathieu form
\cite{mathieu}. The solution to these equation are to be matched at
$U = U^0$. The energy distance relation can be derived using the null
geodesic equation and, for non-zero $K$, leads to the same general
structure as what we found for the unsmeared solution\footnote{It is
interesting to note that for $K=0$, the energy converges to a finite
quantity of order $U^0/\sqrt{gN}$ as $U \rightarrow 0$.  This suggests
that the theory is empty at energies below this scale. In fact, the
field theory dual to this simple background seems to exhibit mass-gap
and confinement.}.

This conclusion seems rather robust and independent of the details of
the derivation.  If we insist that (a) $E(U)$ be given by (\ref{eua})
down to $U=U^0$, (b) that $E(U)$ to be given by (\ref{euc}) deep in
the IR, and (c) that $E(U)$ be a monotonic function, then $E(U)$ must
remain more or less flat in the region $\sqrt{K/N} U^0 < U < U^0$.

We seem to have no choice but to conclude that although $U^0$ and
$(K/N)^{1/4} U^0$ appear to be well separated in the radial scale,
from the point of view of the energy via the ``energy-distance''
relation, the two scales are not separated at all. In fact, they are
all at the Higgs scale.  It seems quite remarkable that the detailed
structure of full D3-brane metric is encoded in the ``fine structure''
of the Higgsed SYM at around the Higgs scale.  With a sufficiently
powerful spectrum analyzer and with a lot of patience, all of the
details of say the absorption and Hawking emission by the D3-brane
including all Mathieu \cite{mathieu} and quantum corrections is in
principle reconstructible from the gauge theory. From a practical
point of view, however, the fact that so much information is crammed
in such a narrow scale in energy suggests that it would be extremely
difficult to isolate and extract information of interest from the full
theory.

\section*{Acknowledgements}

I would like to thank Steve Gubser, Sunny Itzhaki, Finn Larsen, and
Joe Polchinski for illuminating discussions. This work was supported
in part by the National Science Foundation under Grant
No. PHY94-07194.

\begingroup\raggedright\endgroup


\begin{thebibliography}{10}

\bibitem{adscft}
J.~Maldacena, ``The Large N limit of superconformal field theories and
  supergravity,'' {\em Adv. Theor. Math. Phys.} {\bf 2} (1998) 231,
  \href{http://xxx.lanl.gov/abs/hep-th/9711200}{{\tt hep-th/9711200}}.

\bibitem{absorb1}
I.~R. Klebanov, ``World volume approach to absorption by nondilatonic branes,''
  {\em Nucl. Phys.} {\bf B496} (1997) 231,
  \href{http://xxx.lanl.gov/abs/hep-th/9702076}{{\tt hep-th/9702076}}.

\bibitem{absorb2}
S.~S. Gubser, I.~R. Klebanov, and A.~A. Tseytlin, ``String theory and classical
  absorption by three-branes,'' {\em Nucl. Phys.} {\bf B499} (1997) 217,
  \href{http://xxx.lanl.gov/abs/hep-th/9703040}{{\tt hep-th/9703040}}.

\bibitem{absorb3}
S.~S. Gubser and I.~R. Klebanov, ``Absorption by branes and Schwinger terms in
  the world volume theory,'' {\em Phys. Lett.} {\bf B413} (1997) 41--48,
  \href{http://xxx.lanl.gov/abs/hep-th/9708005}{{\tt hep-th/9708005}}.

\bibitem{gkp}
S.~S. Gubser, I.~R. Klebanov, and A.~M. Polyakov, ``Gauge theory correlators
  from noncritical string theory,'' {\em Phys. Lett.} {\bf B428} (1998) 105,
  \href{http://xxx.lanl.gov/abs/hep-th/9802109}{{\tt hep-th/9802109}}.

\bibitem{wittenads}
E.~Witten, ``Anti-de Sitter space and holography,'' {\em Adv. Theor. Math.
  Phys.} {\bf 2} (1998) 253, \href{http://xxx.lanl.gov/abs/hep-th/9802150}{{\tt
  hep-th/9802150}}.

\bibitem{thooftholo}
G.~'t~Hooft, ``Dimensional reduction in quantum gravity,''
  \href{http://xxx.lanl.gov/abs/gr-qc/9310026}{{\tt gr-qc/9310026}}.

\bibitem{sussholo}
L.~Susskind, ``The World as a hologram,'' {\em J. Math. Phys.} {\bf 36} (1995)
  6377--6396, \href{http://xxx.lanl.gov/abs/hep-th/9409089}{{\tt
  hep-th/9409089}}.

\bibitem{susswitt}
L.~Susskind and E.~Witten, ``The Holographic bound in anti-de Sitter space,''
  \href{http://xxx.lanl.gov/abs/hep-th/9805114}{{\tt hep-th/9805114}}.

\bibitem{GHKK}
S.~S. Gubser, A.~Hashimoto, I.~R. Klebanov, and M.~Krasnitz, ``Scalar
  absorption and the breaking of the world volume conformal invariance,'' {\em
  Nucl. Phys.} {\bf B526} (1998) 393,
  \href{http://xxx.lanl.gov/abs/hep-th/9803023}{{\tt hep-th/9803023}}.

\bibitem{witstr98}
E.~Witten, talk presented at Strings 98 Conference,
  http://www.itp.ucsb.edu/online/strings98/witten/.

\bibitem{joeS}
J.~Polchinski, ``S matrices from AdS space-time,''
  \href{http://xxx.lanl.gov/abs/hep-th/9901076}{{\tt hep-th/9901076}}.

\bibitem{sussflat}
L.~Susskind, ``Holography in the flat space limit,''
  \href{http://xxx.lanl.gov/abs/hep-th/9901079}{{\tt hep-th/9901079}}.

\bibitem{IMSY}
N.~Itzhaki, J.~M. Maldacena, J.~Sonnenschein, and S.~Yankielowicz,
  ``Supergravity and the large N limit of theories with sixteen supercharges,''
  {\em Phys. Rev.} {\bf D58} (1998) 046004,
  \href{http://xxx.lanl.gov/abs/hep-th/9802042}{{\tt hep-th/9802042}}.

\bibitem{rg1}
L.~Girardello, M.~Petrini, M.~Porrati, and A.~Zaffaroni, ``Novel local CFT and
  exact results on perturbations of N=4 superYang Mills from AdS dynamics,''
  {\em JHEP} {\bf 12} (1998) 022,
  \href{http://xxx.lanl.gov/abs/hep-th/9810126}{{\tt hep-th/9810126}}.

\bibitem{rg2}
J.~Distler and F.~Zamora, ``Nonsupersymmetric conformal field theories from
  stable anti-de Sitter spaces,''
  \href{http://xxx.lanl.gov/abs/hep-th/9810206}{{\tt hep-th/9810206}}.

\bibitem{rg3}
M.~Porrati and A.~Starinets, ``RG fixed points in supergravity duals of 4-D
  field theory and asymptotically AdS spaces,''
  \href{http://xxx.lanl.gov/abs/hep-th/9903085}{{\tt hep-th/9903085}}.

\bibitem{DougTayl}
M.~R. Douglas and W.~Taylor, ``Branes in the bulk of Anti-de Sitter space,''
  \href{http://xxx.lanl.gov/abs/hep-th/9807225}{{\tt hep-th/9807225}}.

\bibitem{YYWu}
Y.-Y. Wu, ``A Note on AdS/SYM correspondence on the Coulomb branch,''
  \href{http://xxx.lanl.gov/abs/hep-th/9809055}{{\tt hep-th/9809055}}.

\bibitem{PeetPol}
A.~W. Peet and J.~Polchinski, ``UV/IR relations in AdS dynamics,'' {\em Phys.
  Rev.} {\bf D59} (1999) 065006,
  \href{http://xxx.lanl.gov/abs/hep-th/9809022}{{\tt hep-th/9809022}}.

\bibitem{JuanWilson}
J.~Maldacena, ``Wilson loops in large N field theories,'' {\em Phys. Rev.
  Lett.} {\bf 80} (1998) 4859,
  \href{http://xxx.lanl.gov/abs/hep-th/9803002}{{\tt hep-th/9803002}}.

\bibitem{BISY}
A.~Brandhuber, N.~Itzhaki, J.~Sonnenschein, and S.~Yankielowicz, ``Wilson
  loops, confinement, and phase transitions in large N gauge theories from
  supergravity,'' {\em JHEP} {\bf 06} (1998) 001,
  \href{http://xxx.lanl.gov/abs/hep-th/9803263}{{\tt hep-th/9803263}}.

\bibitem{MinWarn}
J.~A. Minahan and N.~P. Warner, ``Quark potentials in the Higgs phase of large
  N supersymmetric Yang-Mills theories,'' {\em JHEP} {\bf 06} (1998) 005,
  \href{http://xxx.lanl.gov/abs/hep-th/9805104}{{\tt hep-th/9805104}}.

\bibitem{GarySunny}
G.~T. Horowitz and N.~Itzhaki, ``Black holes, shock waves, and causality in the
  AdS/CFT correspondence,'' \href{http://xxx.lanl.gov/abs/hep-th/9901012}{{\tt
  hep-th/9901012}}.

\bibitem{KLT}
P.~Kraus, F.~Larsen, and S.~P. Trivedi, ``The Coulomb branch of gauge theory
  from rotating branes,'' \href{http://xxx.lanl.gov/abs/hep-th/9811120}{{\tt
  hep-th/9811120}}.

\bibitem{mathieu}
S.~S. Gubser and A.~Hashimoto, ``Exact absorption probabilities for the
  D3-brane,'' \href{http://xxx.lanl.gov/abs/hep-th/9805140}{{\tt
  hep-th/9805140}}.

\end{thebibliography}
\end{document}